\documentclass[twocolumn,secnumarabic, amssymb, nobibnotes, aps, pra, superscriptaddress]{revtex4-2}

\usepackage[hidelinks]{hyperref}
\usepackage[dvipsnames]{xcolor}
\usepackage{graphicx}
\usepackage{physics}
\usepackage{amsmath}
\usepackage{amsthm}
\usepackage{algorithm2e}
\usepackage{enumitem}
\usepackage{comment}
\usepackage{siunitx}
\usepackage{booktabs}
\newtheorem{theorem*}{Statement}
\DeclareMathOperator*{\argmin}{\arg\!\min} 
\begin{document}


\title{Characterization of coherent errors in gate layers with robustness to Pauli noise}


\newcommand{\fraunhoferIAF}{\affiliation{Fraunhofer Institute for Applied Solid State Physics IAF, Tullastr. 72, 79108 Freiburg, Germany}}
\newcommand{\ETHZ}{\affiliation{Institute for Quantum Electronics, ETH Z\"{u}rich, Otto-Stern-Weg 1, 8093 Z\"{u}rich, Switzerland}}
\newcommand{\HyQ}{\affiliation{Center for Hybrid Quantum Networks (Hy-Q), The Niels Bohr Institute, University of Copenhagen, Blegdamsvej 17, DK-2100 Copenhagen Ø, Denmark}}

\author{Noah Kaufmann}
\email{noah.kaufmann@nbi.ku.dk}
\ETHZ
\HyQ
\author{Ivan Rojkov}
\ETHZ
\author{Florentin Reiter}
\email{freiter@phys.ethz.ch}
\ETHZ
\fraunhoferIAF


\begin{abstract}
    Characterization of quantum devices generates insights into their sources of disturbances. State-of-the-art characterization protocols often focus on incoherent noise and eliminate coherent errors when using Pauli or Clifford twirling techniques. This approach biases the structure of the effective noise and adds a circuit and sampling overhead. We motivate the extension of an incoherent local Pauli noise model to coherent errors and present a practical characterization protocol for an arbitrary gate layer. Notably, the coherent noise estimation is robust to Pauli noise. We demonstrate our protocol on a superconducting hardware platform and identify the leading coherent errors. To verify the characterized noise structure, we mitigate its coherent and incoherent components using a gate-level coherent noise mitigation scheme in conjunction with probabilistic error cancellation. The proposed characterization procedure opens up possibilities for device calibration, hardware development, and improvement of error mitigation and correction techniques.\vspace{0.9cm}
\end{abstract}

\maketitle

\section{Introduction}

Current quantum computers are highly affected by noise~\cite{NISQ}. The errors originate from interactions of the devices with their environment~\cite{wilen_correlated_2021}, unwanted dynamics between the qubits~\cite{krinner_benchmarking_2020}, or imperfect control signals~\cite{NoiseSources}. Characterization aims to find error sources and quantify their effect on the performance of a computing processor~\cite{nielsen_characterization}. The information obtained from the characterization of a device facilitates the development of a less error-prone hardware platform~\cite{co-design}, the implementation of schemes for correcting specific errors~\cite{EfficientLearningQN}, and the application of noise mitigation protocols~\cite{PEC}. Therefore, noise characterization stands at the core of future technological progress. 

Demarcating noise along the lines of purity conservation leads to the distinction between coherent and incoherent noise~\cite{EstimatingCoherenceOfNoise}.
\textit{Coherent} noise occurs due to systematic, reversible perturbations such as imperfect calibrations, imprecise control signals, or couplings to low-frequency external fields~\cite{muhonen_storing_2014,fogarty_nonexponential_2015,suter_colloquium_2016,ball_role_2016}. Conversely, \textit{incoherent} noise (e.g., bit-flip errors) is stochastic and typically arises from insufficient isolation of the system from its environment~\cite{suter_colloquium_2016}, causing a nonreversible loss of information. Since controllability and isolation of a system are generally opposing goals in hardware design~\cite{NoiseSources, tradeoff_cc}, distinguishing the structure of the limiting noise is critical to the advancement of quantum computing platforms.

Numerous characterization techniques have been developed to identify and quantify coherent and incoherent noise in quantum processes. Full-process tomography~\cite{FQPT, FQPT2} and gate-set tomography~\cite{GST1, SPAM_FQPT} are common techniques for obtaining a general representation of a physical operation. Although they provide a full representation, they lack discrimination of the noise processes from the actual operation. Furthermore, as they involve the reconstruction of full-system density matrices, without further assumptions, they are experimentally intractable on devices with more than a few qubits due to exponential scaling of the number of operations with the size of the system's Hilbert space and the resource-intensive postprocessing~\cite{haffner_scalable_2005}. While some protocols achieve scalable noise characterization using principles of randomized benchmarking~\cite{RB_Emerson, RB_Knill, RB_Magesan}, they are often limited to a few heuristic metrics that do not provide enough insights for existing noise calibration, mitigation, or correction techniques~\cite{CT_detection}. Characterization approaches that exceed this limitation while still being scalable exist for incoherent noise~\cite{PEC_IBM, EfficientLearningQN, harper2021fast}. These \textit{noise reconstruction} protocols~\cite{tradeoff} introduce a model and estimate its parameters by fitting it to the real process. They aim to quantify noise of a gate layer, where a restricted noise correlation length is the central structural assumption. However, those models do not portray the actual noise process completely, as the protocols require the application of probabilistic \textit{twirling} schemes~\cite{kern_quantum_2005,randomized_compiling_2,RB_Emerson,emerson_symmetrized_2007,moussa_practical_2012,Twirling} that convert coherent errors into incoherent ones. Currently, no similar approach to noise reconstruction protocols exists that is sensitive to coherent errors.

In this paper, we introduce a noise model combining coherent and incoherent components. We devise a protocol to characterize the coherent part of the model as a series of unitary transformations, interpretable as qubit rotations. The estimation is independent of the incoherent model components introduced, hardware-agnostic, and scalable to larger systems, assuming a low noise-correlation length. To demonstrate our method, we apply it to characterize a gate layer on a seven-qubit superconducting circuit device provided by IBM Quantum (IBMq)~\cite{IBMq}. This allows us to determine the most significant coherent single- and two-qubit errors and their behavior over time. Leveraging these estimation results, we mitigate the coherent errors through a gate-level approach and combine it with probabilistic error cancelation (PEC)~\cite{PEC,li_efficient_2017} in order to verify that our model captures the dominating noise terms. Our work paves the way for advancements in hardware platform development, quantum error correction and mitigation schemes, and the design of noise-aware algorithms.


\section{Model}

An ideal noiseless operation on a quantum computer can be expressed by a \textit{unitary channel} $\mathcal{U}_I(\rho) = U_I \rho U_I^\dag$, for some unitary matrix $U_I$. A simple model for the noisy implementation of this operation on real hardware $\mathcal{E}_{\mathcal{P}}$ consists of the concatenation of a noise modeling \textit{Pauli channel} $\mathcal{P}$ and the ideal unitary,
\begin{equation}
    \label{eq:PModel}
    \mathcal{E}_{\mathcal{P}} \left( \rho \right) = \mathcal{P}\!\left(\,\mathcal{U}_I(\rho)\right) = \sum_i{p(i) P_i U_I \rho U_I^\dag P_i},
\end{equation}
where the Pauli error rates $p(i)$ form a probability distribution over the $n$-qubit Pauli operators $P_i\in\mathbb{P}^{\otimes n}=\{I,X,Y,Z\}^{\otimes n}$. Despite the substantial expressibility restrictions of the model by only considering Pauli operators, a wide range of unital incoherent noise, including dephasing and depolarization, can be described by Pauli channels. In this framework, an $n$-qubit noisy operation is described at most by $4^n$ Pauli error rates. To reduce this scaling from exponential to polynomial, we follow the widely used and experimentally motivated approach of only regarding \textit{locally} correlated noise, that is, considering $n$-qubit Pauli operators $P_i$ that act nontrivially on $l$ qubits only~\cite{PEC_IBM, EfficientLearningQN}. These $l$~qubits are selected with respect to the qubits' connectivity in the quantum computing architecture. This work focuses on a correlation length $l=2$.

Pauli noise models are appealing for different physical and practical reasons. Depolarization and dephasing often dominate the qubit systems' interaction with the environment~\cite{depolarizing_reaction, sanders}. Quantum error correction also leads to noise better approximated by Pauli channels~\cite{QEC_decoheres, QEC_decoheres_2, huang_performance_2019}. Additionally, Pauli channels are interesting for theoretical work as they can be efficiently simulated classically~\cite{Gottesmann_Knill}. Most importantly, however, any noise can be projected onto a Pauli channel by applying Pauli twirling~\cite{kern_quantum_2005,randomized_compiling_2,RB_Emerson,emerson_symmetrized_2007,moussa_practical_2012, Twirling}. If $\mathcal{U}_I$ is a Clifford operation Pauli twirling can be achieved by stochastically introducing Pauli gates before and after the ideal operation. Enforcing the structure of the model given in Eq.~\eqref{eq:PModel} by twirling allows the mitigation of any unknown general noise channel with schemes tailored to Pauli noise, such as PEC. Despite the sampling and circuit overhead due to the probabilistic nature of PEC and twirling~\cite{kern_quantum_2005,randomized_compiling_2,PEC_IBM}, this mitigation approach is well suited for near-term devices because it is implementable with low-depth circuits and relies on Pauli gates~\cite{PEC, PEC_IBM}.

\begin{figure}
    \centering   \includegraphics[width=\columnwidth]{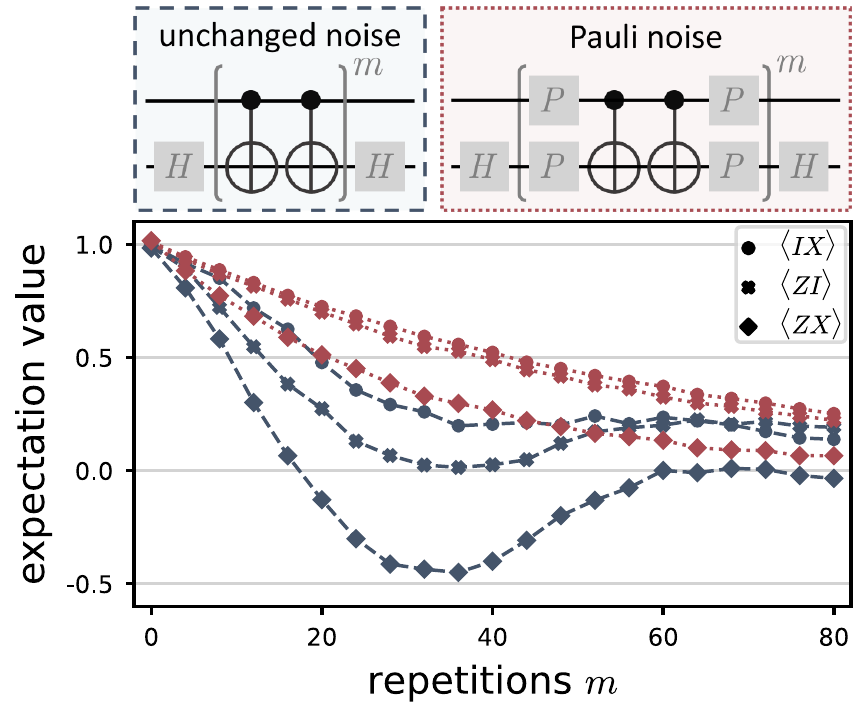}   
    \caption{Noise examination on identity circuit. The data are obtained on \texttt{ibm\_lagos}, and the expectation value of the three two-qubit Pauli operators $P_{IX}, P_{ZI}, P_{ZX}$ is plotted against the number of noisy identities $m$ implemented by two $C\!X$ gates. In the demonstration, we prepare and measure the system in the state $\ket{0+}$ using one Hadamard gate $H$. The blue dashed curves correspond to the unchanged noise. The dotted red curves are obtained using a Pauli twirling scheme that projects the noise to Pauli noise (see Eq.~\eqref{eq:PModel}). Each circuit is measured $4096$ times and the twirled results are averaged over $32$ different twirling gates $P$. A readout mitigation scheme~\cite{readout_mitigation_ibm} is applied in postprocessing based on the reported readout error rates of IBM Quantum.  The results for different input states and superconducting hardware platforms available on IBMq are similar.\\
    }
    \label{fig:osc}
\end{figure}

However, in the context of characterization where the goal is to examine the physical noise sources of a system, altering the noise structure using twirling is often not desired, as it obscures the distinction between coherent and incoherent noise. To illustrate the transformative effect of Pauli twirling and motivate the extension of the Pauli noise model in Eq.~\eqref{eq:PModel} for characterization purposes, we perform an \textit{echo experiment}, shown in Fig.~\ref{fig:osc} (similar to Ref.~\cite{randomization}), on an IBMq processor. The circuits consist of preparing the state $\ket{0+}$, an even number of $C\!X$ gates, and measuring the evolved state in the same Pauli basis as that in which it was prepared. As two $C\!X$ gates correspond to the identity, in the noiseless case all measured expectation values are independent of the number of repetitions and equal to $1$. In the exclusive presence of Pauli noise, purely exponential decay of the expectation values, characteristic of incoherent error processes, is expected~\cite{RB_Emerson,RB_Knill,RB_Magesan,randomized_compiling_2,PauliFrameRandomization}. This behavior (red dotted curves) can be observed by adding Pauli twirling to the circuits. In the absence of twirling (blue dashed curves), oscillations appear, a feature that exceeds the expressibility of the Pauli noise model in Eq.~\eqref{eq:PModel}.

\begin{figure}
    \centering  \includegraphics[width=\columnwidth]{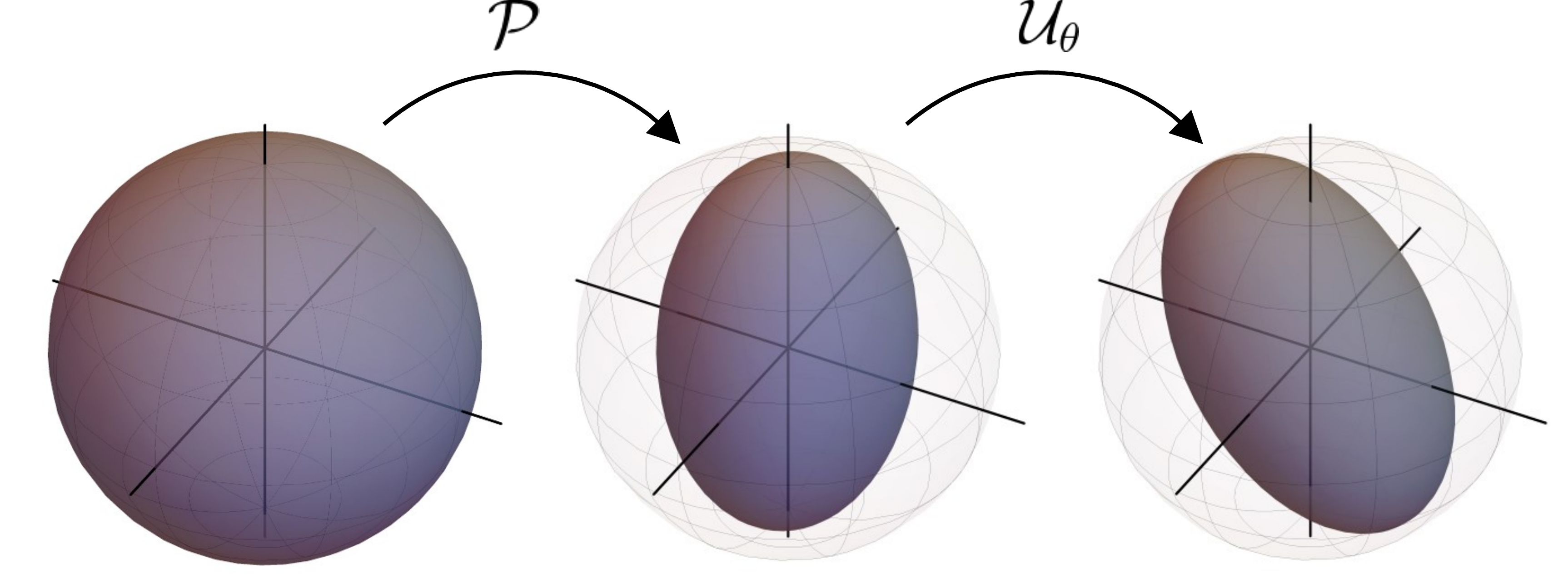}   
    \caption{Bloch sphere transformations. Image of the Bloch sphere on the left under the action of a Pauli channel $\mathcal{P}$ (middle), and an additional unitary transformation $\mathcal{U}_\theta$ (right).}
    \label{fig:bloch_speres}
\end{figure}

The leading cause for the oscillations in Fig.~\ref{fig:osc} is coherent noise. We extend the Pauli noise model in Eq.~\eqref{eq:PModel} with a unitary channel $\mathcal{U}_\theta$ (see Fig.~\ref{fig:bloch_speres}) intended to explain the oscillatory behavior observed in Fig.~\ref{fig:osc} 
\begin{equation}
    \label{eq:CModel}
    \mathcal{E}_{\mathcal{P}, \theta} \left( \rho \right) = \mathcal{U}_\theta\!\left(\,\mathcal{E}_{\mathcal{P}}\!\left( \rho \right) \right) =
    U_\theta \sum_i{p(i) P_i U_I \rho U_I^\dag P_i U_\theta^\dag}
\end{equation}
where $U_\theta$ is a unitary operator parameterized by a Hermitian matrix $H_\theta$ as ${U_\theta = e^{-iH_\theta}}$. This matrix can be expressed in the Pauli basis as $H_\theta = \sum_k \theta_k P_k$, with real coefficients $\theta_k$ and Pauli matrices $P_k$. Assuming small noise, i.e., $U_\theta$ is close to identity, we can rewrite the coherent noise as $U_\theta \approx \prod_k \exp\!\left( -i \theta_k P_k \right)$, which is a valid approximation up to $\mathcal{O}(\theta^2)$. Each product term reads
\begin{equation} \label{eq:small_angle_approx}
    \exp\!\left( -i \theta_k P_k \right) = \cos\left({\theta_k}\right) \mathbb{I} - i \sin\left(\theta_k\right) P_k,
\end{equation}
and represents in the Bloch sphere picture a rotation around the $k$ axis by the angle $2\,\theta_k$, see~Fig.~\ref{fig:bloch_speres}. To obtain a scalable model, we consider, as for the Pauli noise model in Eq.~\eqref{eq:PModel}, locally correlated noise operators, namely single- and two-qubit rotations among neighboring qubits according to their connectivity in the hardware. Therefore, the \textit{coherently rotated Pauli noise model} defined in Eq.~\eqref{eq:CModel} consists of nine two-qubit rotations for each neighboring qubit pair and three single-qubit rotations for each qubit, leading to a magnitude of these rotations of $\lvert\theta\rvert=15$.

Note that this model is not universal. Nonunital noise~\cite{extendedPauli}, such as decay, surpasses its expressibility. In Ref.~\cite{carignan2019polar}, the authors analyze the decomposition of general noise into its coherent and incoherent components, investigating how each part contributes to a bound on the total average process fidelity. They show that, for systems with more than one qubit, a physical unitary channel may not fully account for the channel’s rotational components. Pauli noise does not introduce such rotational components, due to their diagonal Pauli-Liouville representation~\cite{Kimmel2014}. Consequently, when the system's incoherent noise is assumed to be Pauli noise, such as dephasing, only the presence of additional coherent noise can lead to oscillatory features in the noise profile. In the discourse regarding the influence of coherent noise on error correcting schemes and the fault tolerance threshold, similar noise models to Eq.~\eqref{eq:CModel} are utilized in Refs.~\cite{coherent_qecbeginners, CoherentNoisemodelZ1, coherent_model_3}. However, those models are limited to specific types of Pauli and coherent noise, and the works do not focus on the characterization of a noisy process but on analyzing the properties of a process with the respective noise model.

\section{Characterization}
\label{sec:characterization}

We now present our novel protocol for characterizing all the parameters of the coherently rotated Pauli noise model introduced in Eq.~\eqref{eq:CModel}. The protocol relies on the assumption that the amplitude of both coherent and incoherent noise is small.

We express the coherent noise characterization protocol in the \textit{Pauli transfer matrix} (PTM) formalism~\cite{greenbaum_introduction_2015}. As the Pauli matrices constitute a complete basis of the operator space, a $n$-qubit state $\rho$ can be vectorized $\left|\rho \right) \in \mathbb{R}^{4^n}$ in the Pauli basis, with the $i$th element $\left|\rho \right)_i$ being their expectation values ${\left|\rho \right)_i = \frac{1}{2^n} \mathrm{tr} \left( {P_i \rho} \right)}$. These $4^n$ expectation values can be estimated from $3^n$ measurements of $\rho$ in different Pauli bases~\cite{riofrio_experimental_2017,huang_efficient_2021,ClSh_Huang,EfficientLearningQN, schwemmer2014experimental, toth2010permutationally}. For a single qubit, this state representation is known as the Bloch vector~\cite{Bloch_Original}. In this formalism, a channel $\mathcal{E}$ is represented by a matrix $T \in \mathbb{R}^{4^n} \times \mathbb{R}^{4^n}$ that acts on the vectorized states according to $\left| \mathcal{E}\!\left( \rho \right) \right) = T \left| \rho \right)$. The matrix $T$ is called the PTM, with elements $T_{ij} = \mathrm{tr} \left( P_i \,\mathcal{E}\!\left( P_j \right) \right)$. In the context of the coherently rotated Pauli noise model, the diagonal of the PTM matrix encompasses information about the incoherent Pauli noise channel mixed with the second-order terms of the coherent noise. The first-order terms of the coherent noise are contained in the off-diagonal elements without any contribution from Pauli noise.

Before addressing the characterization protocol applicable to a quantum process modeled by the coherently rotated Pauli noise model in Eq.~\eqref{eq:CModel}, we present the estimation if $U_\theta$ for a two-qubit process described by $\mathcal{E}_{\theta}\!\left( \rho \right) = U_\theta \rho U_\theta^\dag$, meaning that we disregard both the Pauli noise $\mathcal{N}$ and the ideal unitary $\mathcal{U}_I$ channels. The Pauli transfer matrix $T_{\theta}$ corresponding to $\mathcal{E}_{\theta}$ is a square matrix of order $16$, parameterized by $15$ single- and two-qubit rotation angles. While the first row and column are trivial and independent of $\theta$, in a first-order approximation in $\theta$ (see App.~\ref{app:second_order_approximation} for a second-order approximation), the other elements of this matrix are given by
\begin{equation} 
    \label{eq:PMT_unitary_rotation}
    \left( T_{\theta} \right)_{ab} \approx \delta_{ab} - \frac{i}{4} \sum_{k}\theta_k \mathrm{tr} \left( \left[P_a, P_b \right] P_k \right).
\end{equation}
By estimating the vector representation of a set of states $\{\rho_i\}$ and their propagation $\{\mathcal{E}_{\theta}\!\left( \rho_i \right)\}$ we get an estimate of $\theta$ by solving the minimization problem
\begin{equation} 
    \label{eq:minimization}
    \hat{\theta} = \argmin_\phi \sum_i \big\lVert \left| \mathcal{E}_{\theta}\!\left( \rho_i \right) \right) -  T_{\phi} \left|\rho_i \right) \big\rVert^2.
\end{equation}
Note that, compared to full-process tomography on two qubits, the number of real parameters is reduced from $240$ to $15$, and for a system with $l$ qubits from $\mathcal{O} \left( 16^l\right)$ to $\mathcal{O} \left( 4^l\right)$. In App.~\ref{app:closed_form} we provide the closed-form solution of the minimization in Eq.~\ref{eq:minimization} for a first-order approximation in $\theta$.

The strategy of the characterization protocol for the entire $n$-qubit system is to apply this estimation procedure to every two-qubit subsystem involved according to their connectivity in the hardware. Three points must be considered when moving from the simple case discussed to the model given in Eq.~\eqref{eq:CModel}: first, the action of the ideal unitary $\mathcal{U}_I$; second, the presence of the Pauli noise~$\mathcal{N}$; and lastly, the impact of two-qubit noise on qubits that are outside of the considered subsystem.

The effect of the ideal unitary channel $\mathcal{U}_I$ can be incorporated in postprocessing after estimating the state $\left| \rho \right)$. For an arbitrary $U_I$, this procedure requires a full-state tomography of the system~\cite{Gottesmann_Knill}. To preserve scalability, we restrict to operations, separable into multiple single- and two-qubit unitaries $U_I = U_{1} \otimes U_{2} \otimes ... \otimes U_{l}$. This requirement is fulfilled, for example, in the characterization of so-called \textit{gate layers}, that is, $n$-qubit operations consisting of single- and two-qubit gates that can be executed simultaneously.

We address the presence of the Pauli noise channel $\mathcal{N}$ by assuming that both the incoherent and coherent noises are relatively small. In the scenario where ${\max_{ijk}\, \lvert p(i) - p(j) \rvert \, \lvert \theta_k \rvert \ll \, 1}$, in which $p(i)$ and $\theta_k$ are respectively the Pauli and the coherent error rates in Eq.~\eqref{eq:CModel}, we can consider the Pauli channel to approximately commute with the coherent rotations. Then we can show that to first order in $\theta$ the estimation procedure presented is not disturbed by the presence of an additional Pauli channel (see App.~\ref{app:PauliCoh}). Intuitively this is visible when illustrating the action of the two types of noise for a single qubit on a Bloch sphere. While coherent noise corresponds to a rotation, Pauli noise leads to a shrinkage of the sphere (see Fig.~\ref{fig:bloch_speres}) which does not alter the orientation of the sphere poles and, therefore, does not interfere with the estimation of unitary rotations. 

Lastly, when focusing exclusively on two-qubit subsystems, one has to address the correlated multiqubit noise between a qubit in the subsystem of interest and the external ones. To handle this problem in the estimation of $\left|\rho \right)$ and $\left|\mathcal{E}_{\theta}\!\left( \rho \right)\right)$, we prepare the surrounding qubits in different basis states for every single run. This technique isolates the two desired qubits and randomizes the effect of the noise introduced outside the subsystem of interest, similar to twirling schemes.

The protocol for the estimation of coherent noise for a two-qubit system is as follows (for more than two qubits see App.~\ref{app:char_steps}):
\begin{enumerate}
\itemsep0em 
    \item Choose a positive integer $j$.
    \item Initialize the two-qubit system in $\ket{0}^{\otimes 2}$.
    \item Prepare the system in the eigenstate of a random two-qubit Pauli operator $P$ by applying to each qubit $H$, $H X$, $S H$, $S H X$, $X$ or $\mathbb{I}$, obtaining a state $\ket{\rho}$, with $P \ket{\rho} = \pm \ket{\rho}$.
    \item Apply $j$ repetitions of the unitary $U$ to the circuit, resulting in $\mathcal{E}_{\theta}^j\!\left( \rho_j \right)$.
    \item Perform state tomography and analytically reverse the action of the ideal unitary $\mathcal{U}_I$. This step results in the state description of $\left| \mathcal{U}_I^\dagger \, \mathcal{E}_{\theta}^j\!\left( \rho_j  \right) \right)$ and $\left| \mathcal{E}_{\theta}^j\!\left( \rho_j \right)\right)$.
    \item Repeat steps $4$ and $5$ for $j+1$.
    \item Repeat steps $1$ to $6$ $L$ times.
    \item Aggregate all pairs $\left\{\left| \mathcal{E}_{\theta}^{j}\!\left( \rho_j \right)\right), \left|\mathcal{U}^\dagger_I \, \mathcal{E}_{\theta}^{j+1}\!\left( \rho_j \right)\right)\right\}$ for all prepared states and repetitions $j$ and carry out the minimization of Eq.~\eqref{eq:minimization}.
\end{enumerate}

\begin{figure}
    \centering  \includegraphics[width=\columnwidth]{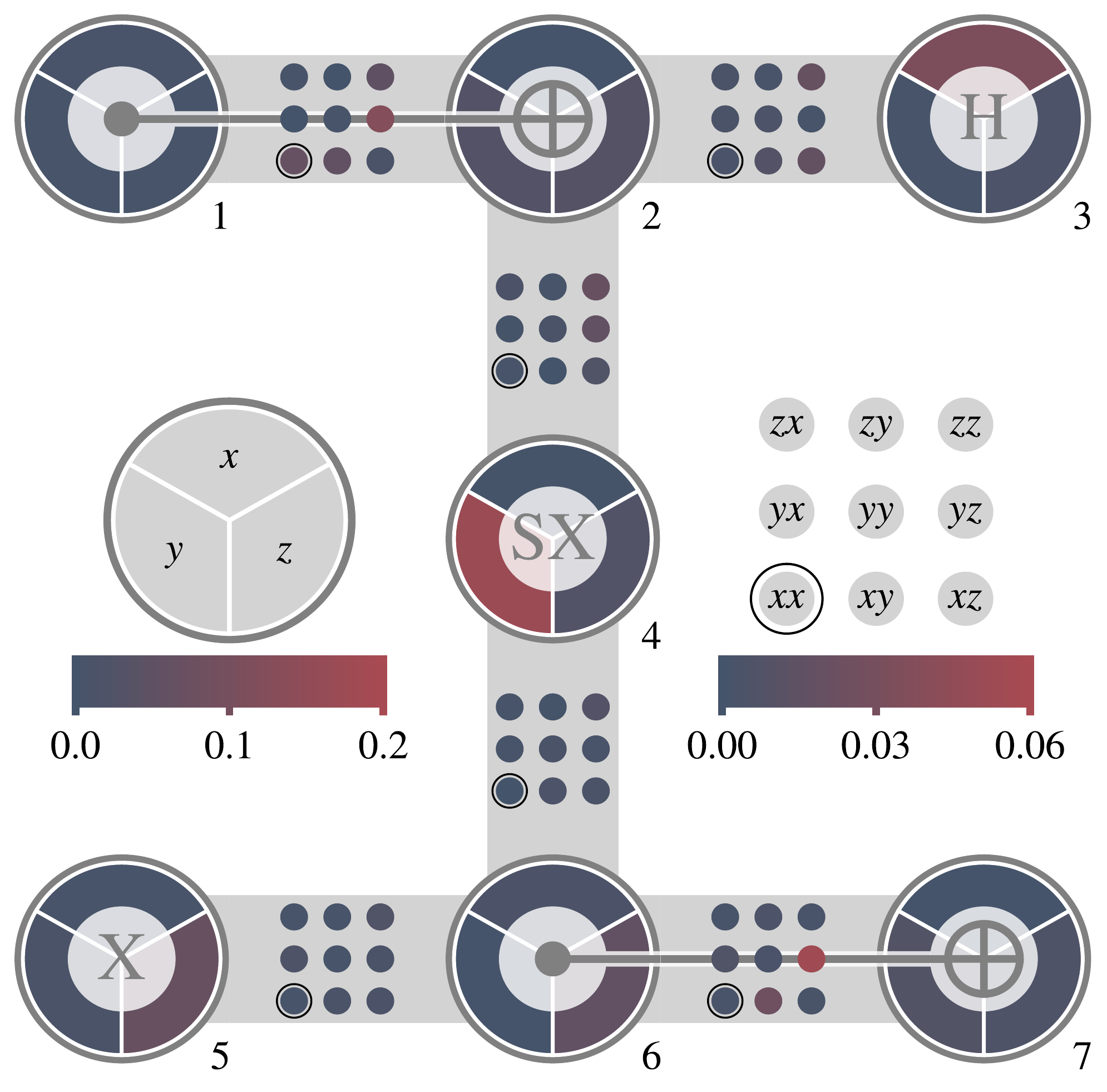}   
    \caption{Characterization of the coherent noise introduced by executing a gate layer on the seven-qubit IBMq Falcon processor \texttt{ibm\_lagos}~\cite{IBMq} (specifications in App.~\ref{app:system_specificatons}). The gate layer is drawn on top of the connectivity tree (qubit connectivity in the quantum computing architecture), where each big circle marks a different qubit. The gate layer consists of two $C\!X$ gates between qubits $1$ and $2$, and qubits $6$ and $7$, a Hadamard ($H$) gate on qubit 3, a $\sqrt{X}$ ($S\!X$) gate on qubit $4$, and an $X$ gate on qubit $5$. The three pieces inside the qubit representing circles represent the single-qubit coherent errors, and the nine small circles between two qubits display the two-qubit coherent cross-talk errors. All errors are measured in radians and the absolute value is plotted.}
    \label{fig:gate_layer_charcter}
\end{figure}

We now use the above protocol to characterize the coherent noise induced by an arbitrary gate layer executed on the IBMq Falcon processor \texttt{ibm\_lagos}. The system specifications are listed in App.~\ref{app:system_specificatons}. The result is summarized in Fig.~\ref{fig:gate_layer_charcter}, where we show the absolute values of the angles of the characterized single- and two-qubit coherent errors for an arbitrary gate layer. The layer is composed of two $C \! X$ gates and three single-qubit gates, namely $H$, $\sqrt{X}$, and $X$. Considering the connectivity of the seven qubits in the hardware and our assumption that the noise is only locally correlated, we prepare $216$ initial states, all being product states of $\ket{0}$, $\ket{1}$, $\ket{+}=(\ket{0}+\ket{1})/\sqrt{2}$, $\ket{-}=(\ket{0}-\ket{1})/\sqrt{2}$, $\ket{i}=(\ket{0}+i\ket{1})/\sqrt{2}$ and $\ket{-i}=(\ket{0}-i\ket{1})/\sqrt{2}$. Each two-qubit subsystem is prepared in $36$ states, and for each of those states, the environment is randomized with six additional states. We execute $0$ to $3$ repetitions of the layer, amounting to four circuits per preparation. Finally, bypassing the effect of the unitary gate layer $\mathcal{U}_I$ in the postprocessing necessitates performing state tomography on up to three-qubit subsystems, leading to $3^3$ Pauli measurements per circuit. Overall, we run $216 \cdot 4 \cdot 27 = 23328$ different circuits with $128$ shots each, leading to a total running time in the system of about $30$ minutes. A more detailed description of the procedure is given in App.~\ref{app:char_steps}.

\begin{figure*}
    \centering  \includegraphics[width=\textwidth]{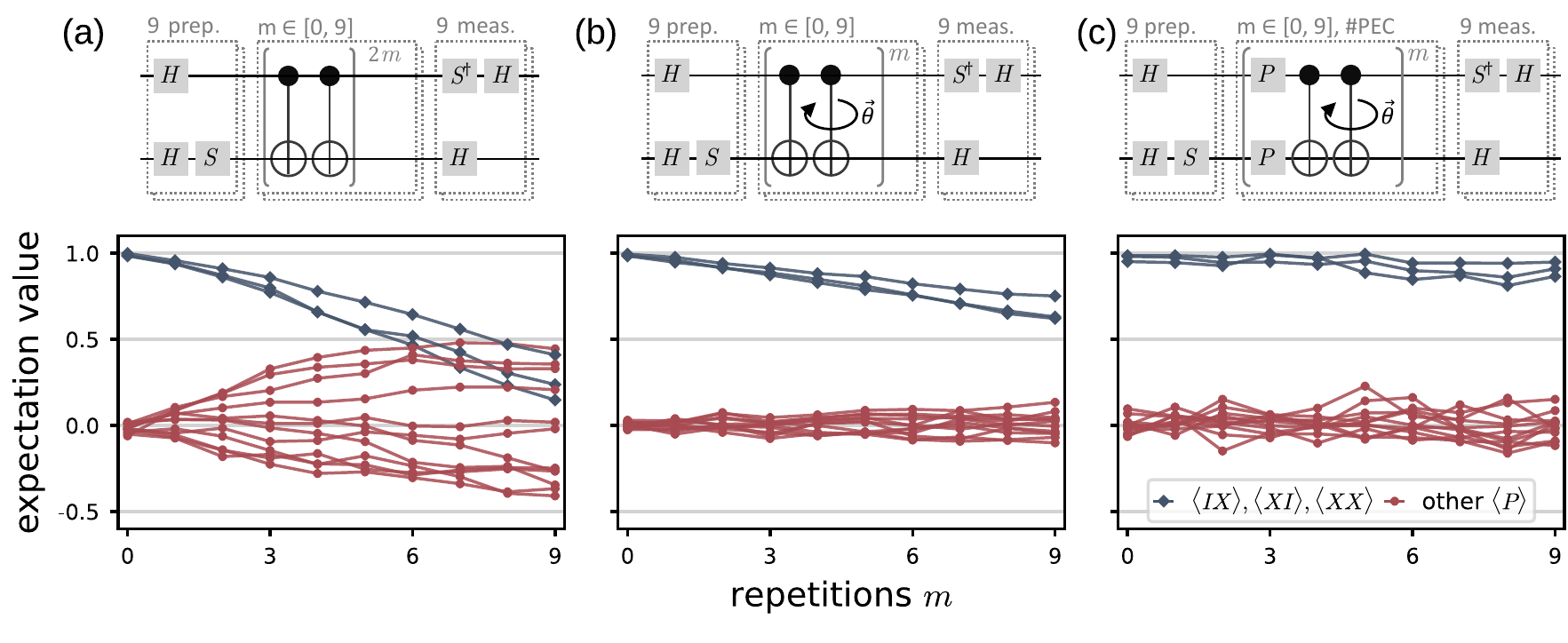} 
    \caption{Noise mitigation according to the coherently rotated Pauli noise model in Eq.~\eqref{eq:CModel} on \texttt{ibm\_lagos}. The plot shows the estimated expectation values of all 15 two-qubit Pauli operators (excluding $\mathbb{I} \otimes \mathbb{I}$) for the initial state $\ket{++}$ after $m$ repetitions of a noisy identity (a) without mitigation, (b) with coherent error mitigation based on data of (a), and (c) with probabilistic error cancelation based on data of (b) and coherent error mitigation. $\ket{++}$ is an eigenstate of the operators  $P_{I\!X}$, $P_{X\!I}$, and $P_{X\!X}$, in the noiseless case, and their expectation values are $1$. The expectation values of the other 12 Pauli operators (red) are $0$. The circuit diagrams contain the preparation of each qubit in one of the states ${\ket{0}, \ket{+}, \ket{i}}$, by applying Hadamard ($H$) and/or phase ($S$) gates. The evolved states are measured in nine bases to estimate the $15$ Pauli expectation values. In (c), specific Pauli operators ($P$) are inserted between the circuit elements according to the PEC method.}
    \label{fig:coherent}
\end{figure*}

The IBMq platforms used report readout errors of the order of 1\%, caused by relaxation, imperfect coupling to the readout resonator, and signal amplification errors~\cite{readout_ibm_report}. To counter the influence of these errors, we apply a postprocessing readout mitigation scheme in all demonstrations~\cite{readout_mitigation_ibm}. While our protocol is not resistant to general state preparation and measurement errors, probabilistic errors that can be modeled by a Pauli channel do not disturb the characterization for the same reasons that the coherent noise characterization is robust against Pauli noise. This is described in detail in App.~\ref{app:PauliCoh}. Furthermore, systematic, coherent, preparation errors also do not affect the protocol, as can be seen in Eq.~\ref{eq:minimization} where the estimation does not depend on information about the intended prepared state.

From Fig.~\ref{fig:gate_layer_charcter}, we conclude that the most significant two-qubit coherent errors occur between qubits $1$ and $2$ and between qubits $6$ and $7$. Those are associated with the Pauli $P_{Y\!Z}$ error. This result is expected, as those are the pairs of qubits on which the $C\!X$ gates are acting. This Pauli error has already been noted in the analysis of the echoed cross-resonance gate used to realize the $C\!X$ gate~\cite{sundaresan_reducing_2020}. The predominant coherent errors are the single-qubit ones. We note no strong bias towards a specific single-qubit rotation error, and those errors could be related to the hardware implementation of each single-qubit gate.

We conduct the characterization of the same device and gate layer as in Fig.~\ref{fig:gate_layer_charcter} three times, with time differences between the runs being 6 hours and 19 days. The results available in App.~\ref{app:coherent_drift} show that the leading coherent errors in the system do not change over time. The most significant single-qubit coherent errors change by no more than $10\%$ and the two-qubit coherent rotation errors by maximally $0.018$ radians. This result suggests that the platform suffers from systematic coherent errors not influenced by the daily recalibration of the system.


\section{Mitigation} 
In principle, the characterization results in Fig.~\ref{fig:gate_layer_charcter} may originate from overfitting the nonideal process to our model. Indeed, it is not known whether the coherently rotated Pauli noise model in Eq.~\eqref{eq:CModel} is expressive enough to resemble the main characteristics of the probed IBMq process. To verify that the model captures the dominant noise contributions of a process on the probed device, we examine whether the output of the presented protocol facilitates significant noise mitigation

Returning to the unmitigated circuits in Fig.~\ref{fig:osc}, our goal is to suppress the observed oscillatory behavior in the echo experiment by correcting the characterized coherent errors at the circuit level.  In the case of a successful correction, according to our model, one would expect a purely exponential decay of the expectation values with the number of process repetitions due to the Pauli noise. From this exponential decay, we can then characterize the Pauli error rates~\cite{EfficientLearningQN, PEC_IBM} and mitigate these noise components using PEC~\cite{PEC}. If the final measurement statistics after coherent error mitigation and PEC closely follow the ideal statistics, we have demonstrated that the proposed model is physical, and the characterization protocol provides valuable insights.

Any unitary channel is \textit{reversible}. Therefore, in our model, the effect of the coherent noise can be canceled by single- and two-qubit rotations with opposite rotation angles to those characterized. We introduce a parameterized circuit element (see Fig.~\ref{fig:c_element}) capable of representing all required rotations while corresponding to the identity if all parameters are set to zero (see App.~\ref{app:coherent_mitigation}). We use this circuit element to run echo experiments for three different situations: unmitigated; coherent-error mitigated; and coherent-error mitigated combined with PEC. The results are presented in Fig.~\ref{fig:coherent}. Fig.~\ref{fig:coherent}(a) and (b) were obtained with 2048 shots per circuit, whereas for Fig.~\ref{fig:coherent}(c), an ensemble of 280 altered circuits with 512 shots each was executed per experiment.  

For the unmitigated case in Fig.~\ref{fig:coherent}(a), we see two main characteristics. Firstly, the expectation values of the 12 Pauli operators that should ideally be $0$ are spreading. Secondly, the expectation values that ideally equal $1$, meaning that the states are eigenstates of the corresponding Pauli, decay.

Applying our coherent noise characterization protocol to the data of Fig.~\ref{fig:coherent}(a) leads to a first estimate of the rotation angles of the coherent errors $\hat{\theta}_0$. To further improve the characterization, we run the characterization scheme with $\vec{\theta}=\hat{\theta}_0$ and obtain $\hat{\theta}_1$. This iterative approach is expected to improve the characterization because it compensates for violations of the assumed commutation of single- and two-qubit rotation errors. Fig.~\ref{fig:coherent}(b) shows the result of the final mitigation iteration with the estimate $\vec{\theta}=\hat{\theta}_1$ using the gate-based mitigation circuit from Fig.~\ref{fig:c_element}. The observed changes in the angles between the first and the last iteration were about an order of magnitude smaller than the biggest coherent errors and thus support the validity of the approximation taken. In Fig.~\ref{fig:coherent}(b), the mitigation effect is clearly visible, as the spreading behavior of the expectation values that are ideally $0$ is largely suppressed. The expectation values of the Pauli operators, of which the prepared state is a $+1$ eigenstate, are higher in the mitigated case compared to the unmitigated estimation. However, the coherent noise mitigation does not retrieve their ideal-case expectation value of $1$, as according to our model the statistic is still affected by Pauli noise.

\begin{figure}
    \centering  \includegraphics[width=\columnwidth]{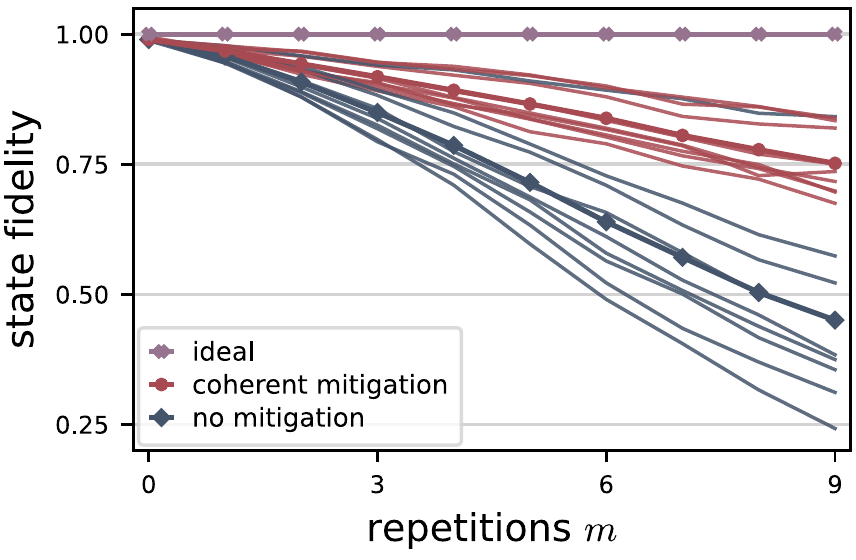}
    \caption{Influence of coherent error mitigation on fidelity measurement between the ideal state and states evolved by the noisy identity. From the data of Fig.~\ref{fig:mit_all}, the density matrix of the evolved state is estimated for each number of repetitions of the noisy identity and each of the nine prepared states. The plot shows the fidelity between these state estimates and the ideally prepared state. The thin lines represent the evolution of the fidelity with the number of repetitions of the noisy identity for each state, and the thick lines are the average of the nine thin lines.}
    \label{fig:fidelities}
\end{figure}

To show that the noise examined follows the coherently rotated Pauli noise model in Eq.~\eqref{eq:CModel}, we estimate the Pauli error rates from the mitigated curves of Fig.~\ref{fig:coherent}(b) and apply PEC to mitigate the Pauli noise. Note that, different from the above-presented coherent noise correction, PEC is a mitigation technique that does not correct individual circuits deterministically (as Pauli channels are not reversible) but allows noise-free estimation of expectation values from circuit ensembles. It is important to emphasize, that compared to previous works demonstrating PEC~\cite{endo_practical_2018,PEC_IBM}, we do not use Pauli twirling and thus avoid the associated circuit and sampling overhead. The Pauli eigenvalues of the Pauli noise channel are estimated by fitting exponential curves to the expectation values that are supposed to be 1 in the ideal case. The Pauli error rates can then be found by a Walsh-Hadamard transform~\cite{EfficientLearningQN}. We follow the PEC protocol of \citet{PEC_IBM} to mitigate the incoherent noise. This involves applying the nonphysical inverse of a Pauli channel by running an ensemble of unitary processes composing this inverse. Those results are weighted according to the Pauli error rates and combined in postprocessing. The results of the echo experiments with coherent error mitigation and PEC are displayed in Fig.~\ref{fig:coherent}(c). Compared to Fig.~\ref{fig:coherent}(b), the data are noisier, which we suspect to be due to the PEC sampling overhead. 

The measurement statistics obtained clearly resemble the main features of the ideal measurement statistics. Therefore, we can conclude that our noise model explains the leading noise terms in the system examined and that the characterization protocol presented generates valid physical insights into the noise. The data of other input states is shown in App.~\ref{app:complementary_mitigation_data}. Similar characterization and mitigation results were obtained on \texttt{ibm\_jakarta}.

Based on the data from Fig~\ref{fig:coherent}(a) and (b), we can reconstruct the density matrices for the nine prepared initial states at every repetition of the noisy identity. We then estimate their fidelity with their ideal initial state. The standard state fidelity formula is used $F(\sigma, \rho) = \mathrm{tr} \left( \sqrt{\sqrt{\rho}\sigma\sqrt{\rho}} \right)^2$. We observe in Fig.~\ref{fig:fidelities} that the mitigation decreases the average infidelity by about a factor of 2.

\section{Generalization of the protocol}

We motivated our coherently rotated Pauli noise model and devised the protocol for its characterization in a manner agnostic toward the choice of the qubit-based quantum computing architecture. The same procedure can be employed to characterize coherent errors in superconducting devices as we demonstrated in this work, but also in trapped ions, neutral atoms, or any other platforms, as it only requires the preparation of elementary Pauli states without the need for entanglement.

Furthermore, we can readily extend the proposed model to qudit-based quantum processors, which are also susceptible to coherent errors~\cite{blok_quantum_2021,morvan_qutrit_2021,goss_high-fidelity_2022,cao_efficient_2022,goss_extending_2023}. In these systems, Pauli $X$ and $Z$ operators are generalized as ${X_d\ket{s}=\ket{s+1\,\,\mathrm{mod}\,\,d}}$ and ${Z_d\ket{s}=\omega^s\ket{s}}$, respectively. Here, $\ket{s}$ represents a qudit computational state with $s\in\{1,2,\ldots,d\}$ and $\omega=\exp(i2\pi/d)$. Consequently, the Pauli noise model for an $n$-qudit system can be expressed similarly to Eq.~\eqref{eq:PModel} but with $P_i\in\{(X_d)^p (Z_d)^q \,\, \vert \,\, q,p \in \{1,2,\ldots,d\}\}^{\otimes n}$ representing the generalized Pauli operators. Since the generalized Pauli operators exhibit similar properties to qubit Pauli gates (such as being unitary 1-designs and normalized by the qudit Clifford group), the Pauli transfer matrix formalism remains valid in the qudit framework. It thus enables the same twirling and characterization techniques employed in the qubit case to be utilized~\cite{blok_quantum_2021,morvan_qutrit_2021,goss_high-fidelity_2022,cao_efficient_2022,goss_extending_2023}. Moreover, this implies that the unitary errors can be effectively modeled by the coherently rotated Pauli noise model described in Eq.~\eqref{eq:CModel} and that the characterization protocol presented in Sec.~\ref{sec:characterization} can be utilized. Additionally, the expansion of the protocol to generalized Pauli operators is also of interest for the characterization of coherent leakage of the qubit. Coherent leakage involves a unitary interaction between the computational subspace and the neglected part of the system's Hilbert space or its environment~\cite{wood2018quantification}. Leakage back to the system introduces noise that is correlated to a past state of the system, introducing memory effects~\cite{wallman2016robust}. By expanding the system description to states in the leakage subspace, it is possible to describe the coherent leakage processes equally to coherent noise within the computational subspace. However, this demands precise control or generalized measurements of the leakage subspace, which must be clearly defined and localized.

Similarly, our characterization protocol can be expanded to encompass the logical level, as logical qubits can be susceptible to unitary noise often caused by coherent errors at the physical level~\cite{gutierrez_errors_2016, greenbaum_modeling_2017}. To extend our protocol, it becomes necessary to prepare and measure logical Pauli basis states. These operations are inherent to various error correcting codes, thereby enabling the generalization of our protocol. 

\section{Conclusion and Outlook}

In this paper, we introduced a characterization technique to estimate coherent errors in noisy quantum devices. An existing Pauli noise model is extended to incorporate coherent errors by adding unitary rotations to the model. We devise a procedure to learn the unknown parameters of this coherently rotated Pauli noise model. The scalability of our approach is affirmed by assuming a limited correlation length of the coherent noise. As the results of the characterization of a physical quantum processor showed, it is feasible to execute the protocol on an actual device. For the device tested, the characterization revealed the leading coherent errors introduced by the probed gate layer. Notably, these characterized coherent errors remained stable over multiple recalibration cycles, suggesting a systematic nature. To verify the validity of our noise model, we mitigated the coherent errors and the characterized Pauli noise. Coherent error mitigation was achieved with a gate-level correction scheme for two-qubit coherent errors. To counter the Pauli noise, we applied probabilistic error cancelation. The combination of the two techniques resulted in almost ideal measurement statistics. 
This finding shows that the model is sufficiently expressible to explain the main noise contributions of a specific process on an actual processor. Moreover, the successful mitigation of the characterized errors confirms the applicability of the developed protocol. Our protocol is hardware agnostic and thus expected to be executable on a variety of hardware platforms. 

Our work offers several promising avenues for future research. The method can enhance mitigation schemes that rely on detailed knowledge of the noise structure. One such approach is optimizing the resilience to coherent noise during circuit transpilation based on the noise insights for each gate or gate layer~\cite{HI_1, HiddenInverse, zhang_hidden_2022, robustnes_bounds, murphy2019controlling, robustnes_bounds}. A clear example of this is the hidden inverses method~\cite{HiddenInverse, zhang_hidden_2022}, which mitigates the impact of coherent errors by compiling circuits in a way that induces destructive interference among the coherent errors of the gates. However, implementing hidden inverses requires knowledge of the physical nature of these errors, which in Ref.~\cite{zhang_hidden_2022} was obtained through gate-set tomography. Integrating our approach into the calibration process can provide valuable information to the compiler, enabling it to leverage the hidden inverses technique while actively inverting the noncorrected coherent errors. Similarly, our technique opens the door to studying and developing noise-aware versions of advanced mitigation techniques such as dynamical decoupling or composite pulses. Moreover, information on coherent errors facilitates Pauli conjugation~\cite{cai_mitigating_2020}. An efficient noise twirling method that limits sampling overhead compared to common Pauli twirling and, therefore, diminishes the reduction of the threshold of error-correcting codes.

Finally, identifying the nature and strength of the errors is also crucial for the fault-tolerant operation of a quantum processor unit at the logical level~\cite{gutierrez_errors_2016,HarmfulCoherentErrors,iyer_small_2018,huang_performance_2019}. While quantum error correction protocols transform local unitary errors into correctable probabilistic ones~\cite{QEC_decoheres, QEC_decoheres_2}, coherent noise can lead to significantly larger worst-case logical errors than only incoherent noise~\cite{wallman_randomized_2014,sanders,EstimatingCoherenceOfNoise,CoherentNoisemodelZ1, gutierrez_errors_2016,greenbaum_modeling_2017,QEC_decoheres_2,coherentNoise,iyer_small_2018,huang_performance_2019}. The knowledge gained from coherent errors, which effectively represents soft information, can then be leveraged to enhance the performance of quantum error-correcting codes at both the decoding and encoding stages. Indeed, while optimal decoders are notoriously hard to find~\cite{iyer_hardness_2015}, recent studies have demonstrated the benefit of soft information in improving existing decoders and overall quantum error correction performance~\cite{xue_repetitive_2020,pattison_improved_2021,raveendran_soft_2022}. At the encoding level, it is possible to design codes to be naturally robust against specific coherent noises, as shown in Refs.~\cite{debroy_optimizing_2021,ouyang_avoiding_2021}. Thus, our protocol has the potential to be a diagnostic tool that could help to decide on the next higher-level encoding.

In the long run, the characterization protocol presented paves the way for the development of improved hardware platforms by identifying the limiting imperfections. Yet, already in the near term it can serve as a tool for various applications that rely on knowledge of the existing noise structure, such as quantum error correction or hardware-aware algorithm design.

\begin{acknowledgments}
    The authors thank Jonathan Home, Elias Zapusek, Roland Matt, Jeremy Flannery, and Luca Huber for helpful comments and discussions throughout the project. 
    This work was supported by the Swiss National Science Foundation (SNSF) through the National Centre of Competence in Research-Quantum Science and Technology (NCCR QSIT) Grant 51NF40–160591. I.R. and F.R. acknowledge financial support by the Swiss National Science Foundation (Ambizione Grant No. PZ00P2$\_$186040). We acknowledge the use of IBM Quantum services for this work. The views expressed are those of the authors, and do not reflect the official policy or position of IBM or the IBM Quantum team.
\end{acknowledgments}

\section*{Data Availability}
The data supporting the findings of this study, including simulation and experimental results, are not publicly available but can be obtained from the corresponding author upon reasonable request.

\appendix
\section*{Appendices}

\section{Second order approximation in $\theta$ for PTM elements}
\label{app:second_order_approximation}
To find the PTM element of $U_\theta$ up to the second-order of $\theta$, we first expand $U_\theta = \Pi_k \text{exp}(-i \theta_k P_k)$ using a second order approximation of Eq.~\eqref{eq:small_angle_approx}:
\begin{equation}
    \begin{aligned}
        U_\theta &\approx \prod_k \left( \left( 1 - \frac{\theta_k^2}{2} \right) \mathbb{I} - i \theta_k P_k \right)\\
        &\approx \frac{2 -\Sigma_k \theta_k^2}{2} \mathbb{I} - i \sum_k \theta_k P_k - \sum_k \sum_{l>k} \theta_k \theta_l P_k P_l,
    \end{aligned}
\end{equation}
We now use this approximation to evaluate the elements $\left[ T_\theta \right]_{ab}$ of the PTM up to the second order of $\theta$:
\begin{equation}
    \begin{aligned}
        \left[ T_\theta \right]_{ab} = \frac{1}{2^n} &\mathrm{tr}\Bigl(P_a U_\theta^\dagger P_b U_\theta \Bigr) \\
        \approx \frac{1}{2^n} \mathrm{tr} \Biggl( P_a &P_b \left(1- \sum_k \theta_k^2\right) -i \sum_k \theta_k \left[P_a, P_b\right] P_k\\
        - &\sum_k \sum_{l>k} \theta_k \theta_l \left(P_a P_l P_k P_b +  P_a P_b P_k P_l \right) \\
        + &\sum_k \sum_l \theta_k \theta_l P_a P_k P_b P_l\Biggr)
    \end{aligned}
\end{equation}
We can split the last term of the equation into the cases where $k>l$, $k<l$ and $k=l$ and simplify to:
\begin{equation}
    \begin{aligned}
        &\sum_k \sum_l \theta_k \theta_l P_a P_k P_b P_l = \sum_k \theta_k^2 P_a P_k P_b P_k\\
        & + \sum_k \sum_{l>k} \theta_k \theta_l \left( P_a P_k P_b P_l + P_a P_l P_b P_k\right) \\
    \end{aligned}
\end{equation}
Using that $P_a P_k P_b P_k = (-1)^{\langle k, b \rangle} P_a P_b$, where $\langle k, b \rangle$ is $0$ if $P_b$ and $P_k$ commute and $1$ otherwise, and the cyclic property of the trace leads to
\begin{equation}
    \begin{aligned}
        \left[ T_\theta \right]_{ab}  \approx \frac{1}{2^n}  \mathrm{tr} \Biggl[ &\left( 1 \!- \! \sum_k \theta_k^2 \left(1\!-\!(-1)^{\langle b, k \rangle}\right) \right) P_a P_b \\
        & - i \sum_k \theta_k \left[P_a, P_b \right] P_k \\
        & + \sum_k \sum_{l>k} \theta_k \theta_l \left[ P_a, P_l \right] \left[ P_b, P_k \right] \Biggr],     
    \end{aligned}
\end{equation}
where all terms of order in $\theta$ greater than 2 are dropped. Finally, we obtain the approximation of the PTM element of $U_\theta$ up to the second order in $\theta$:
\begin{equation}
    \begin{aligned}
        \left[ T_\theta \right]_{ab}  \approx  &\left( 1 - \sum_k 2 \theta_k^2 \langle a, k \rangle \right) \delta_{ab} \\
        & - \frac{i}{2^n} \sum_k \theta_k \mathrm{tr}\left(\left[P_a, P_b \right] P_k\right) \\
        & + \frac{1}{2^n} \sum_k \sum_{l>k} \theta_k \theta_l \mathrm{tr}\left(\left[ P_a, P_l \right] \left[ P_b, P_k \right] \right).   
    \end{aligned}
\end{equation}
Assuming the coherent noise is sufficiently weak, we can simplify this expression to obtain the approximation in Eq.~\eqref{eq:PMT_unitary_rotation}.

\section{Closed form expression for estimation}
\label{app:closed_form}
The right-hand side of Eq.~\eqref{eq:PMT_unitary_rotation} can be written equivalently as
\begin{equation} \label{eq:PMT_unitary_rotation_first order}
        \left( T_{\theta} \right)_{ij} \approx \delta_{ij} - \frac{i}{4} \theta_k \mathrm{tr} \left( \left[P_i, P_j \right] P_k \right) =
    \delta_{ij} + \theta_k\,C_{kij},
\end{equation}
where we use Einstein's notation for the summation over~$k$ and $C$ is a tensor with coefficients $k,i$, and $j$. Applying the channel to an arbitrary state~$\left|\rho \right)$ leads to the equation ${\left|\mathcal{E}_{\theta}\!\left( \rho \right) \right)_i = \left|\rho \right)_i + \theta_k\,C_{kij} \left|\rho \right)_j}$, again applying Einstein's notation to sum over $k$ and $j$. Defining the matrix $(B_{\left|\rho \right)})_{ik} = C_{kij} \left|\rho \right)_j$, the estimation problem boils down to a linear inverse problem of the form ${B_{\left|\rho \right) }\,\theta = \left| \mathcal{E}_{\theta}\!\left( \rho \right) \right) - \left|\rho \right)}$ where the vectors $\left|\rho \right)$ and $\left|\mathcal{E}_{\theta}\!\left( \rho \right)\right)$ can be experimentally estimated and $B_{\left|\rho \right)}$ is a function of $\left|\rho \right)$. However, $B_{\left|\rho \right)}$ is singular and cannot be inverted. To overcome this we must estimate $B_{\left|\rho \right)}$, $\left|\rho \right)$ and $\left|\mathcal{E}_{\theta}\!\left( \rho \right)\right)$ for multiple states $\rho$. Doing this for $N$ states, we formulate the overdetermined system $\Tilde{B} \theta = y$, where $\Tilde{B}$ of size $15N \times 15$ is the concatenation of all matrices $B_{\left|\rho_i \right)}$ and $y$ the concatenation of all vectors $\left| \mathcal{E}_{\theta}\!\left( \rho_i \right) \right) - \left|\rho_i \right)$. Finally, we estimate $\theta$ using the least-squares method, 
\begin{equation}
    \label{eq:closed_form}
    \hat{\theta} = \left( \Tilde{B}^T  \Tilde{B} \right)^{-1} \Tilde{B}^T y \,.
\end{equation}

\begin{figure*}
    \centering  \includegraphics[width=\textwidth]{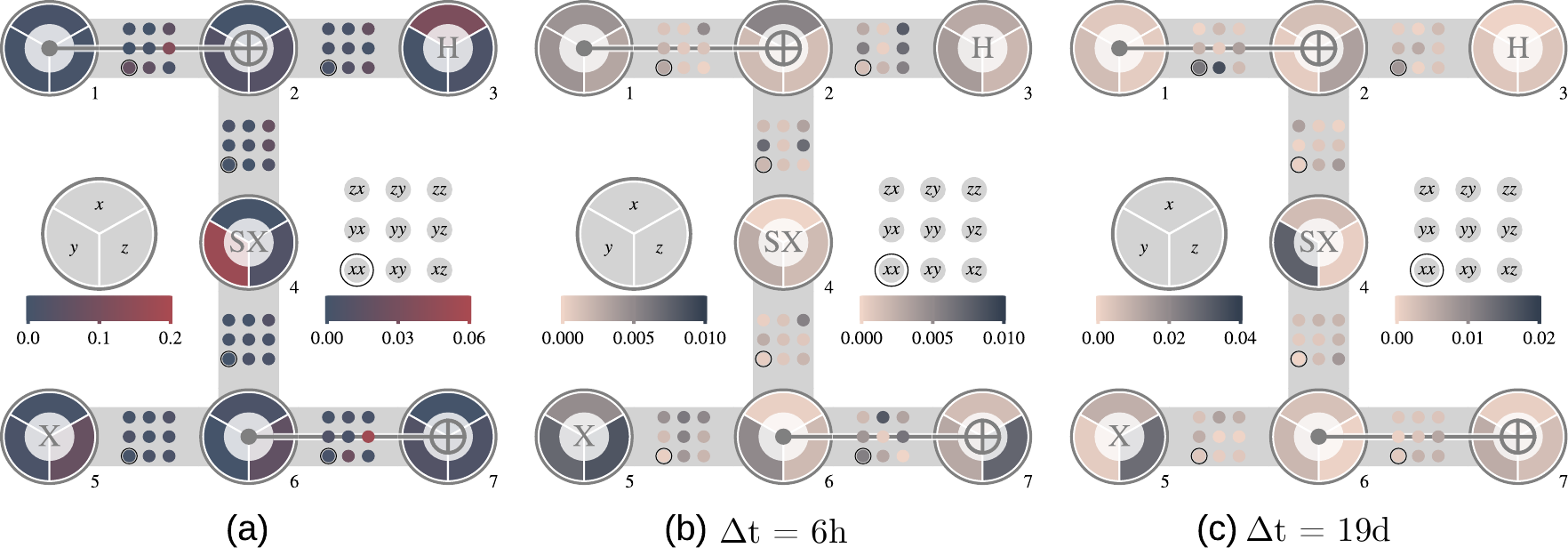}   
    \caption{Coherent noise drift. We repeated the characterization result of Fig.~\ref{fig:gate_layer_charcter} at three different times. Subplot (a) shows the baseline data of Fig.~\ref{fig:gate_layer_charcter}. Subplot (b) displays the observed changes in the coherent error characterization over a time span of 19 days and subplot (c) exhibits the differences over 6 hours. The circuits were executed on the seven-qubit IBM Quantum processor \texttt{ibm\_lagos}~\cite{IBMq} (specifications in App.~\ref{app:system_specificatons}) on September 22, 2022 and October 11, 2022.} 
    \label{fig:gate_layer_charcter_diff}
\end{figure*}

\section{Addressing Pauli noise neglect during coherent noise characterization}
\label{app:PauliCoh}
This section complements the intuitive reasoning of Sec.~\ref{sec:characterization}, on why we can neglect the influence of the Pauli noise channel when characterizing the present coherent noise, with a mathematically more rigorous explanation.

The action of a noisy channel modeled by $\mathcal{E}_{\mathcal{P}, \theta}(\rho)$ as defined in Eq.~\eqref{eq:CModel} can be represented in terms of the vectorized form of the density matrix of $\rho$ and the PTM of the channel. It reads ${\left| \mathcal{E}_{\mathcal{P}, \theta}(\rho) \right) = T_{\mathcal{E}_{\mathcal{P}, \theta}} \left| \rho \right) = T_{U_\theta} T_{\mathcal{P}} T_{U_I} \left| \rho \right)}$, where $T_{U_\theta},\,T_{\mathcal{P}}, \text{ and } T_{U_I}$ are the transfer matrices of the coherent, Pauli, and ideal unitary channels, respectively. As stated in the main text, we discuss the case of $T_{U_I}$ being the identity. Moreover, we assume that $\left| \mathcal{E}_{\mathcal{P}, \theta}(\rho) \right)$ and $\left| \rho \right)$ can be experimentally estimated. 
The estimation of the 15 parameters of $T_{U_\theta}$ can be written as a minimization problem over the estimator~$\hat{T}_{U_\theta}$. 

\begin{theorem*}
Assume $T_{\mathcal{P}}$ and $T_{U_\theta}$ are close to the identity. The minimization of the function
\begin{equation}
    \left\lVert \left| \mathcal{E}_{\mathcal{P}, \theta}(\rho) \right) -  \hat{T}_{U_\theta} T_{\mathcal{P}} \left| \rho \right) \right\rVert ^2
\end{equation} 
over $\hat{T}_{U_\theta}$ is equivalent to the minimization of 
\begin{equation}
    \left\lVert\left| \mathcal{E}_{\mathcal{P}, \theta}(\rho) \right) -  \hat{T}_{U_\theta} \left| \rho \right) \right\rVert ^2\,\,.
\end{equation}
\end{theorem*}

\begin{proof}
A noisy identity process described by the coherent noise model, $ \left| \mathcal{E}_{\mathcal{P}, \theta}(\rho) \right)$ equals $T_{U_\theta} T_{\mathcal{P}} \left| \rho \right)$. We show that $\theta^{\prime} = \theta$ minimizes the expression $\lVert T_{\mathcal{P}} T_{U_\theta} \left| \rho \right) -  T_{U_{\theta^{\prime}}} \left| \rho \right) \rVert$ independent of the choice of $\rho$.
    \begin{equation}
    \begin{aligned}  
    &\min_{\theta^ {\prime}}{\left\lVert\left| \mathcal{E}_{\mathcal{P}, \theta}(\rho) \right) -  {T}_{U_{\theta^{\prime}}} \left| \rho \right) \right\rVert^2}  \\
    &=\min_{\theta^ {\prime}}{\left\lVert  T_{U_\theta} T_{\mathcal{P}} \left| \rho \right) -  {T}_{U_{\theta^{\prime}}} \left| \rho \right) \right\rVert ^2}\\
    &=\min_{\theta^ {\prime}}{
    {\left| \rho \right)}^\top\!\left( T_\mathcal{P}^\top T_{U_\theta}^\top - {T}_{U_{\theta^{\prime}}}^\top \right)\! 
    \Big( T_{U_\theta} T_{\mathcal{P}} - {T}_{U_{\theta^{\prime}}} \Big) \left| \rho \right)}\\
    &=\min_{\theta^{\prime}}
    {\left( \rho \right|} \left( 
    T_\mathcal{P}^\top T_{U_\theta}^\top T_{U_\theta} T_\mathcal{P} - 
    T_{U_{\theta^{\prime}}}^\top T_{U_\theta} T_\mathcal{P} - \right.\\
    &\left.\qquad\qquad\quad-T_\mathcal{P} T_{U_\theta}^\top {T}_{U_{\theta^{\prime}}} +  
    {T}_{U_{\theta^{\prime}}}^\top {T}_{U_{\theta^{\prime}}} \right) \left| \rho \right)\,. 
    \end{aligned}
    \end{equation}
Given that the noise channel is close to the identity, we take the approximation that the diagonal matrix $T_\mathcal{P}$ commutes with the matrices $T_{U_\theta}$ and $T_{U_\theta^\prime}$. Furthermore, because of the unitarity of the coherent rotation channel, $T_{U_\theta}^\top {T_{U_\theta}} = \mathbb{I}$ is valid. Then the minimization problem boils down to
\begin{equation} \label{eq:minimization_problem}
\min_{\theta^{\prime}} \left( \rho \right| \left( \mathbb{I}\!+\!T_\mathcal{P}^2\!-\! T_\mathcal{P}\!\left(T_{U_\theta}^\top {T}_{U_{\theta^{\prime}}}\!+\!\left( T_{U_\theta}^\top {T}_{U_{\theta^{\prime}}} \right)^{\!\!\top} \right) \right) \left| \rho \right)\,.
\end{equation}
The off-diagonal part of $T_{U_\theta}^\top T_{U_{\theta^{\prime}}}$ is antisymmetric. Consequently, $T_{U_\theta}^\top {T}_{U_{\theta^{\prime}}} + \left( T_{U_\theta}^\top {T}_{U_{\theta^{\prime}}} \right)^{\!\!\top}$ is diagonal and less than or equal to ${2 ~ \mathbb{I}}$. Thus, all involved matrices are diagonal. From ${1 + T_{\mathcal{P}}^2 \geq 2~T_{\mathcal{P}}}$, it follows that the expression in Eq.~\eqref{eq:minimization_problem} is minimal when ${T_{U_\theta}^\top {T}_{U_{\theta^{\prime}}} + \left( T_{U_\theta}^\top {T}_{U_{\theta^{\prime}}} \right)^{\!\!\top} = 2 ~ \mathbb{I}}$. This equation has a unique solution corresponding to $ T_{U_\theta} = {T}_{U_{\theta^{\prime}}}$, which finally implies $\theta = \theta^\prime$.

\end{proof}

This statement implies that we can optimize for the coherent noise term without considering the present Pauli noise, allowing us to apply the closed-form solution in Eq.~(\ref{eq:closed_form}) or conduct a least-squares optimization method for $\lVert\left| \mathcal{E}_{\mathcal{P}, \theta}(\rho) \right) -  \hat{T}_{U_\theta} \left| \rho \right) \rVert ^2$.

The approximation of the diagonal matrix $T_\mathcal{P}$ commuting with the matrices $T_{U_\theta}$ is limited by the term ${\max_{ijk}\, \lvert p(i) - p(j) \rvert \, \lvert \theta_k \rvert \ll \, 1}$. The first part of this expression quantifies how isotropic the action of the Pauli noise is. For the depolarizing channel, which for a single qubit is illustrated as a uniformly contracting Bloch sphere, $\max_{ij} \lvert p(i) - p(j) \rvert$ is zero. In that case $T_\mathcal{P}$ can be written as $f \cdot \mathbb{I}$, with $f$ being the depolarizing factor, and thus commutes with all other matrices. The second term of the expression, namely $\max_{k} \lvert \theta_k \rvert$, is a measure of the maximal off-diagonal terms in the coherent noise modeling PTM. As the Pauli noise modeling matrix is diagonal, the violation of the commutation is only facilitated by the off-diagonal of $T_{U_\theta}$. The geometric pictures of scaling and rotation help to understand this commutation relation better.

\section{step-by-step circuit generation}
\label{app:char_steps}
 
We aim to provide a step-by-step overview of the circuit construction for gate-layer characterization. Each circuit has three main parts: state preparation, the gate layer of interest, and measurement. Since we prepare and measure the states in a Pauli basis, those two parts only require the local gates $X$, $H$, $S$, and $S^\dagger$. We choose a set of $L$ states to prepare, ensuring that there is no correlation in the preparation of nearby qubits. These states can be selected randomly; however, as described in Sec.~\ref{sec:characterization}, we choose a more structured set of $216$ initial states that are all combinations of the states $\ket{+}$, $\ket{-}$, $\ket{i}$, $\ket{-i}$, $\ket{1}$, and $\ket{0}$, prepared by applying the gates $H$, $H X$, $S H$, $S H X$, $X$, and $\mathbb{I}$ respectively to the state $\ket{0}$.
Next, we apply $j$ and $j + 1$ iterations of the gate-layer, where $j$ is an integer. At the final measurement stage, we need a basis that allows us to reverse the action of the ideal unitary on each subsystem of connected qubits. For every neighboring pair of qubits, $q_i$ and $q_j$, the required measurement basis must encompass the subsystem that includes these two qubits along with all other qubits sharing a two-qubit gate with them. To achieve this, we use a complete Pauli measurement basis for each subsystem, which consists of all possible combinations of measuring each qubit in the $x$, $y$, or $z$ basis.

The general procedure for arbitrary system sizes and $k$-local noise is as follows:
\begin{enumerate}
\itemsep0em
    \item Choose a positive integer $j$.
    \item Initialize the $n$-qubit system in $\ket{0}^{\otimes n}$.
    \item Prepare the system in the eigenstate of a random $n$-qubit Pauli operator $P$ by applying $H$, $H X$, $S H$, $S H X$, $X$, or $\mathbb{I}$ to each qubit, obtaining a state $\ket{\rho}$, with $P \ket{\rho} = \pm \ket{\rho}$.
    \item Apply $j$ repetitions of the unitary $U$ to the circuit, resulting in $\mathcal{E}_{\theta}^j\!\left( \rho_j \right)$
    \item Perform state tomography on each relevant $l$-qubit subsystem. With $k$ being the assumed locality of the considered noise, $l$ must be large enough to analytically reverse the action of the ideal unitary $\mathcal{U}_I$ for each $k$-qubit subsystem. This step results in the state description of $\mathcal{U}_I^\dagger \, \mathcal{E}_{\theta}^j\!\left( \rho_j \right)$ and $\mathcal{E}_{\theta}^j\!\left( \rho_j \right)$ for each $k$-qubit subsystem.
    \item Repeat steps $4$ and $5$ for $j
    +1$.
    \item Repeat steps $1$ to $6$ $L$ times
    \item For each $k$-qubit subsystem, aggregate all pairs $\left\{\left| \mathcal{E}_{\theta}^{j}\!\left( \rho_j \right)\right), \left|\mathcal{U}^\dagger_I \, \mathcal{E}_{\theta}^{j+1}\!\left( \rho_j \right)\right)\right\}$ for all prepared states and repetitions $j$. Carry out the minimization of Eq.~\eqref{eq:minimization}.
\end{enumerate}

\begin{figure*}
    \centering  \includegraphics[width=0.65\textwidth]{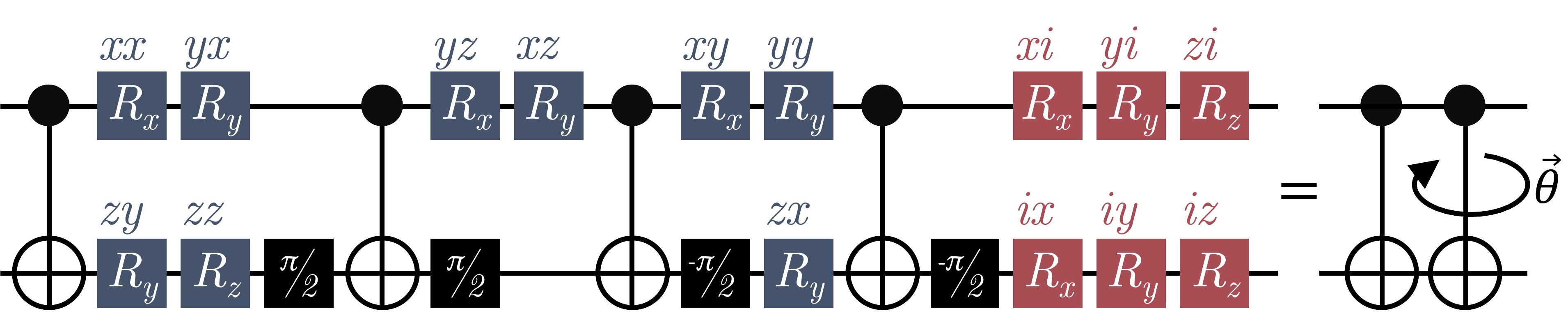}   
    \caption{Coherent noise mitigating circuit element. The circuit element comprises four $C\!X$ gates, four fixed $z$-rotations by $\pi/2$, or $-\pi/2$, and 15 parameterized single-qubit rotations. The first nine rotations (blue) account for different two-qubit coherent errors, and the final six rotations (red) mitigate single-qubit errors. Without the first $C\!X$ gate, the circuit element represents a corrected $C\!X$ gate. We represent the element by the symbol on the right-hand side, where $\vec{\theta}$ corresponds to the angles of the 15 parameterized rotations.}
    \label{fig:c_element}
\end{figure*}

\section{Coherent noise drift} \label{app:coherent_drift}
We conducted the characterization of the gate layer represented in Fig.~\ref{fig:gate_layer_charcter} again after 6 hours and after 19 days. The shifts in the characterization results between those three demonstrations are shown in Fig.~\ref{fig:gate_layer_charcter_diff}. First, we observe that the leading coherent errors in the system did not change over the 19 days. However, as expected, the changes observed for the two demonstrations within 6 hours are significantly smaller than the differences over the complete time span. After 6 hours, the maximum deviation for single-qubit coherent errors was $0.009$ radians and $0.004$ radians on average. The two-qubit errors changed maximally by $0.008$ radians and on average by $0.003$ radians. In the total period (i.e. after 19 days), the most significant single-qubit coherent errors changed by maximally $10\%$ (corresponding to about $0.03$ radians). The two-qubit coherent rotation errors changed by maximally $0.018$ radians. The average deviation was $0.007$ radians for the single-qubit errors and $0.003$ radians for the two-qubit errors. These results suggest that the platform is affected by systematic coherent errors which are not influenced by the hourly and daily recalibrations of the system during which qubits' frequency and readout angle, as well as pulses' amplitude and phase of basic single- and two-qubit gates, are calibrated~\cite{IBMq}.

\begin{figure*}
    \centering  
    \includegraphics[width=\textwidth]{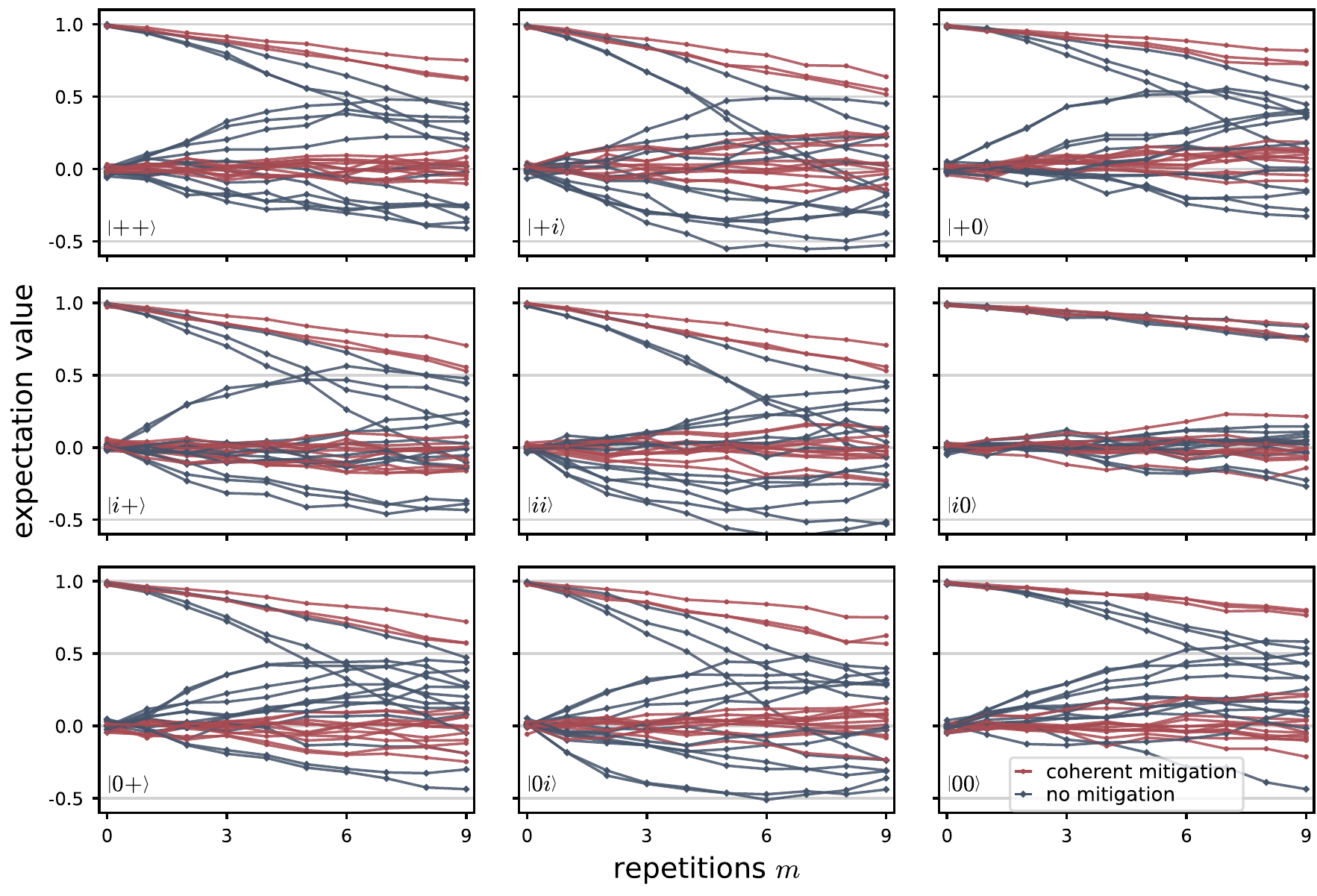}   
    \caption{Complete data set from coherent error mitigation displayed partially in Fig.~\ref{fig:coherent}. The blue curves display the unmitigated results as shown in Fig.~\ref{fig:coherent}~(a). Red shows the curves of the demonstration with coherent mitigation according to Fig.~\ref{fig:coherent}~(b). Each of the 9 different panels exhibits a different initial two-qubit state.}
    \label{fig:mit_all}
\end{figure*}

\section{System specifications}
\label{app:system_specificatons}

Table~\ref{tab:specs_lagos} presents the system specifications of the IBM Quantum Falcon processor \texttt{ibm\_lagos} as provided by IBMq~\cite{IBMq}.

\begin{table}[ht]
\begin{center}
\begin{tabular}{lllllllll}
\toprule
\multicolumn{2}{l}{\textbf{Qubits}} & Q1 & Q2 & Q3 & Q4 & Q5 & Q6& Q7 \\
 & readout-error [$10^{-2}$] & 1.3 & 0.7 & 0.5 & 1.9 & 2.1 & 2.5 & 1.1 \\

 & T1 [\si{\micro\second}] & 68 & 103 & 186 & 167 & 103 & 116 & 108\\
 & T2 [\si{\micro\second}] & 53 & 87 & 101 & 41 & 33 & 89 & 141 \\
 & sx-error [$10^{-4}$] & 2.8 & 2.3 & 1.9 & 1.9 & 2.7 & 1.9 &3.0\\
 & x-error [$10^{-4}$] & 2.8 & 2.3 & 1.9 & 1.9 & 2.7 & 1.9 &3.0\\
\midrule
\multicolumn{2}{l}{\textbf{C\!X-gate}} & 1-2 & 6-7 \\
 & C\!X-error [$10^{-3}$] & $6.8$ & $7.2$\\
 & gate time [\si{\nano\second}] & $305$ & $291$\\
\bottomrule
\end{tabular}
\end{center}
\caption{Specification of the IBM Quantum Falcon processor \texttt{ibm\_lagos} as provided by IBMq~\cite{IBMq} at the time of the characterization on September 22, 2022. The numbering of the qubits refers to the numbers in Fig.~\ref{fig:gate_layer_charcter} and Fig.~\ref{fig:gate_layer_charcter_diff}.}
\label{tab:specs_lagos}
\end{table}

\section{Coherent noise correcting circuit element} \label{app:coherent_mitigation}
To conduct echo experiments that allow for correcting coherent errors we require a parameterized circuit element capable of representing the characterized rotations. Initially, when all coherent noise correction terms are set to zero, the circuit must correspond to the identity circuit. Single-qubit rotations are readily available in most of the leading quantum computing architectures. To invert the two-qubit rotations, an entangling gate is necessary.  We use the $C\!X$ gate as it is a native gate of the chosen platform. By interleaving 15 single-qubit rotations into a structure of four $C\!X$ gates and four fixed single-qubit rotations, as shown in Fig.~\ref{fig:c_element}, a circuit element with 15 parameters is created. Each of the 15 parameters, which we will denote by $\Vec{\theta}$ corresponds to a different single- or two-qubit rotation. The element corresponds to the identity if all parameters are set to zero.

\section{Complementary mitigation data}
\label{app:complementary_mitigation_data}
Fig.~\ref{fig:mit_all} displays the complete data set used to create Fig.~\ref{fig:fidelities}. The influence on the expectation values that ideally are $0$ for all numbers of repetitions is clearly visible for all states, as they disperse much less in the mitigated case. Furthermore, the expectation values of the operators, of which the prepared state is an eigenstate, are closer to the ideal value of $1$ in the mitigated case. While the mitigation did not work equally well for all nine states, the whole data set gives a clear indication of the positive effect of the coherent noise mitigation scheme presented.

\bibliography{aps_main_text}

\end{document}